\documentclass[twocolumn,trackchanges]{aastex62}
\shorttitle{EVOLUTION OF MAGNETIC FIELD WITH FLARE RIBBONS}
\shortauthors{LIU ET AL.}

\def\mathbi#1{\textbf{\em #1}}
\newcommand{\Sdo}{\textit{Solar Dynamics Observatory}}
\newcommand{\sdo}{\textit{SDO}}
\newcommand{\ha}{H$\alpha$}
\newcommand{\hmi}{Helioseismic and Magnetic Imager}
\newcommand{\sm}{$\sim$}
\newcommand{\dg}{$^{\circ}$}
\newcommand{\hinode}{\textit{Hinode}}
\newcommand{\kms}{km~s$^{-1}$}
\newcommand{\GOES}{\textit{Geostationary Operational Environmental Satellite}}

\turnoffeditone

\begin{document}

\title{EVOLUTION OF PHOTOSPHERIC \edit1{VECTOR}\\ MAGNETIC FIELD ASSOCIATED WITH \edit1{MOVING}\\ FLARE RIBBONS AS SEEN BY GST}

\author{Chang Liu}
\affil{Space Weather Research Laboratory, New Jersey Institute of Technology, University Heights, Newark, NJ 07102-1982, USA}
\affil{Big Bear Solar Observatory, New Jersey Institute of Technology, 40386 North Shore Lane, Big Bear City, CA 92314-9672, USA}

\author{Wenda Cao}
\affil{Big Bear Solar Observatory, New Jersey Institute of Technology, 40386 North Shore Lane, Big Bear City, CA 92314-9672, USA}

\author{Jongchul Chae}
\affil{Astronomy Program, Department of Physics and Astronomy, Seoul National University, Seoul 151-747, Korea}

\author{Kwangsu Ahn}
\affil{Big Bear Solar Observatory, New Jersey Institute of Technology, 40386 North Shore Lane, Big Bear City, CA 92314-9672, USA}

\author{Debi Prasad Choudhary}
\affil{Department of Physics and Astronomy, California State University Northridge, Northridge, CA 91330-8268, USA}

\author{Jeongwoo Lee}
\affil{Institute of Space Sciences, Shandong University, Weihai, Shandong 264209, China}
\affil{Space Weather Research Laboratory, New Jersey Institute of Technology, University Heights, Newark, NJ 07102-1982, USA}

\author{Rui Liu}
\affil{CAS Key Laboratory of Geospace Environment, Department of Geophysics and Planetary Sciences,\\University of Science and Technology of China, Hefei 230026, China}

\author{Na Deng}
\affil{Space Weather Research Laboratory, New Jersey Institute of Technology, University Heights, Newark, NJ 07102-1982, USA}

\author{Jiasheng Wang}
\affil{Space Weather Research Laboratory, New Jersey Institute of Technology, University Heights, Newark, NJ 07102-1982, USA}

\author{Haimin Wang}
\affil{Space Weather Research Laboratory, New Jersey Institute of Technology, University Heights, Newark, NJ 07102-1982, USA}
\affil{Big Bear Solar Observatory, New Jersey Institute of Technology, 40386 North Shore Lane, Big Bear City, CA 92314-9672, USA}

\begin{abstract}
The photospheric response to solar flares, also known as coronal back reaction, is often observed as sudden flare-induced changes in vector magnetic field and sunspot motions. However, \edit1{it remains obscure whether evolving flare ribbons, the flare signature closest to the photosphere, are accompanied by changes in vector magnetic field therein}. Here we explore the relationship between the dynamics of flare ribbons in the chromosphere and variations of magnetic fields in the underlying photosphere, \edit1{using } high-resolution off-band \ha\ images and near-infrared vector magnetograms of the M6.5 flare on 2015 June 22 observed with the 1.6~m Goode Solar Telescope. We find that changes of photospheric fields occur at the arrival of the flare ribbon front, thus \edit1{propagating analogously } to flare ribbons. In general, the horizontal field increases and the field lines become more inclined to the surface. When ribbons sweep through regions that undergo a rotational motion, the fields transiently turn more vertical with decreased horizontal field and inclination \edit1{angle}, and then restore and/or become more horizontal than before the ribbon arrival. The ribbon \edit1{propagation decelerates near } the sunspot rotation center, where the vertical field becomes permanently enhanced. Similar magnetic field \edit1{changes } are discernible in magnetograms from the \hmi\ (HMI), and \edit1{an inward collapse of coronal magnetic fields is inferred from } the time sequence of non-linear force-free field models extrapolated from HMI magnetograms. \edit1{We conclude } that photospheric fields respond nearly instantaneously to \edit1{magnetic reconnection in the corona}.
\end{abstract}


\section{INTRODUCTION}\label{introduction}
It is widely believed that the structural evolution and dynamics of the solar photosphere (e.g., magnetic flux emergence and shearing motion) can build up free magnetic energy in the corona that powers flares and coronal mass ejections (CMEs; \citealt{priest02}). The reconfiguration of coronal magnetic field due to energy release is the focus of almost all models of flares/CMEs, which generally do not consider the restructuring of magnetic and flow fields in the dense photosphere partially due to the often assumed line-tying effect \citep{raadu72}. Nonetheless, observational evidences of rapid (in minutes), significant, and permanent photospheric structural changes apparently as a response to flare/CME occurrences have been accumulated over the past 25 years from both ground- and space-based instruments (see e.g., \citealt{wang15} for a recent review). These include stepwise changes of line-of-sight (LOS) and vector magnetic fields \citep[e.g.,][]{wang92,sudol05,wang10,liu12,sun12,petrie12,liu14,song16,sun17,castellanos18}, morphological changes of sunspot penumbrae \citep[e.g.,][]{wang04a,liu05,deng05,xuz16,xuz17}, changes of photospheric flow field \citep[e.g.,][]{tan09,deng11,wangs14,wangjs18}, and sunspot displacement and rotations \citep{anwar93,liu10b,wangs14,liu16,bi16,bi17,xuz17}. Although it is sometimes challenging to disentangle the cause-and-effect relationship between flare/CME processes and photospheric structural changes, studying this topic can shed new insights into the photosphere-corona coupling under the context of energy and momentum transportation in the flare-related phenomena, and help advance and constrain flare/CME models.

The aforementioned various aspects of photospheric evolution closely associated with flares/CMEs were largely studied separately. It might be possible that they can be accommodated by the back reaction of coronal restructuring on the photosphere and interior \citep{hudson08}. In this scenario, the coronal magnetic field would contract inward due to magnetic energy release \citep{hudson00}, and the central photospheric field vectors may be loosely expected to tilt toward the surface (i.e., becoming more horizontal) as a result of this contraction. Such a magnetic field change would correspond to a Lorentz-force change that is exerted at and below the photosphere \citep{hudson08,fisher12,petrie14}. Furthermore, the inward collapse of coronal field might also be accompanied by an upward turning of fields in the peripheral regions \citep{liu05}. These are well in line with observations of flare-induced contraction of coronal loops \citep[e.g.,][]{Liur+implosion09,liur+wang09,liur+wang10,liur12b,gosain12,simoes13,wangj18}, and with photospheric observations that flaring sites usually exhibit an enhancement of horizontal magnetic field $B_h$ and penumbral structure at the center, surrounded by regions of weakened $B_h$ and penumbrae; also, \edit1{the resulting} Lorentz-force change seems to be able to drive the observed surface flows and sunspot motions (see references above). It should be noted that although the overall magnetic field in three dimension (3D) must become more potential after the release of magnetic energy, the near-surface field could become more stressed after flares/CMEs \citep[e.g.,][]{jing08,liu12}.

It is worth noting that due to resolution limitation imposed by data, a majority of previous studies rely on the comparative analysis of pre- and postflare structures. Meanwhile, this approach avoids the concern that heating from flare emissions change spectral line profiles, leading to \textit{transient} anomaly in the magnetic field measurement \citep[e.g.,][]{patterson81,zirin81b,qiu03,maurya12,sun17}. For flare-related \textit{permanent} magnetic field changes, the most prominent one could be the irreversible strengthening of $B_h$ in regions around central flaring PILs and between double flare ribbons. This has been corroborated by results from not only observations but also MHD modeling \citep[e.g.,][]{li11,inoue15,inoue18}. However, there are only rare reports about permanent changes of photospheric magnetic and flow field in association with the spatial and temporal evolution of flare emissions, specifically, flare ribbons. Using LOS magnetograms from the \edit1{Global Oscillation Network Group}, \citet{sudol05} pointed out in several events that the step-like LOS field change appears to propagate at a speed similar to those of ribbons. A propagating motion of $B_h$ enhancement across the flaring region in a major flare event was also noticed by \citet{sun17} using vector magnetic field data from the \hmi\ \citep[HMI;][]{schou12} on board the \Sdo\ (\sdo). Importantly, higher resolution data at both the chromospheric and photospheric levels are needed to fully exploit the association between flare ribbon motions and magnetic/flow field changes, which could provide major clues to the origin of flare-related restructuring on the surface.

Recently, based on chromospheric \ha\ and photospheric TiO images at unprecedented resolution obtained with the 1.6~m Goode Solar Telescope \citep[GST;][]{goode10,cao10,goode12,varsik14} at Big Bear Solar Observatory (BBSO), \citet{liu16} discovered that a sunspot experiences a differential rotation, where the moving front corresponds to a flare ribbon that moves across the sunspot during the 2015 June 22 M6.5 flare event (SOL2015-06-22T18:23) in NOAA active region (AR) 12371. This finding implies that the surface rotation is directly linked to the magnetic reconnection process in the corona \citep{aulanier16}. Naturally, this revives the question of whether the photospheric magnetic field would change permanently as ribbons sweep by. Motivated by our observation, \citet{wheatland18} presented a theoretical model in which this kind of flare-ribbon-related photospheric \edit1{change results} from a downward propagating shear Alfv{\'e}n wave from the coronal reconnection region. Another natural question is whether the velocity $u$ of ribbon propagation would be affected concurrently by the possible field change, since under a simplified two-dimensional magnetic reconnection model, $u$ is correlated with the vertical field $B_z$ on the surface as $u=E/B_z$, where $E$ is the electric field strength in the reconnecting current sheet \citep{forbes84}.

Several other works have also studied this 2015 June 22 M6.5 flare from various perspectives. Mainly using data from BBSO/GST's Visible Imaging Spectrometer (VIS) and Near InfraRed Imaging Spectropolarimeter (NIRIS; \citealt{cao12}), \citet{wang17} reported small preflare brightenings near magnetic channels that may be precursors to the event onset. With nonlinear force-free field (NLFFF) modeling, \citet{Awasthi18} revealed that the initial magnetic reconnection may occur within a multiple flux rope system. \citet{jing17} observed a propagating brightening in the flare decay phase, which may be linked to a slipping-type reconnection. More relevant to the present study, \citet{wangjs18} analyzed GST TiO and HMI observations and found flare-related enhanced penumbral and shear flows as well as $B_h$ around the PIL, which could be attributed to the coronal back reaction. Using HMI observations and NLFFF models, \citet{bi17} presented that the main sunspots on either side of the PIL rotate clockwise during the flaring period, when coronal fields are found to contract significantly. In addition, with NIRIS data \citet{deng17} studied magnetic field property and flare-related evolution of umbral fine structures, and \citet{xu18} showed a transient rotation of surface field vectors seemingly associated with one flare ribbon. Related discussions will be given below.

In this paper, we further investigate the 2015 June 22 M6.5 flare event by comparatively studying high spatiotemporal resolution VIS chromospheric \ha\ off-band images and NIRIS photospheric near-infrared vector magnetograms from BBSO/GST. These state-of-the-art observations are essential for achieving our goal of scrutinizing the intimate relationship between the motion of flare ribbons and possible \textit{permanent} changes of the local vector field, \edit1{which was not studied before}. Special attention is paid to $B_h$, which is the component exhibiting the most clear flare-related changes \citep[e.g.,][]{wang10,wang15,fisher12}. Concerning the aforementioned flare-produced transient magnetic anomaly, we note that the contamination of NIRIS polarimetry from flare emissions was claimed not to be present in this event, as no significant changes are detected in NIRIS intensity profiles \citep[][\edit1{also see the Appendix and Figure~\ref{stokes}}]{xu18}. Moreover, we mainly concern ourselves with permanent magnetic field changes associated with the flare. For the purposes of data validation and results corroboration, HMI vector magnetograms are analyzed as well. In order to examine the evolution of 3D magnetic field above the flaring AR, we also build a time sequence of NLFFF extrapolation models based on HMI data. The plan of this paper is as follows. In Section~\ref{observation}, we first introduce observations and data processing procedures. In Section~\ref{results}, we describe results derived from analyses of observations and magnetic field models, and remark on their implications. More details of structural evolution can be seen in the accompanying animations. In Section~\ref{summary}, we summarize major findings and discuss the results.

\section{OBSERVATIONS AND DATA PROCESSING} \label{observation}
BBSO/GST employs a combination of a high-order adaptive optics system with 308 subapertures \citep{shumko14} and the post-facto speckle-masking image reconstruction technique \citep{woger08}. During \sm16:50--23:00~UT on 2015 June 22, GST makes observations of the then near-disk-center (8$^{\circ}$W, 12$^{\circ}$N) NOAA AR 12371 and achieves diffraction-limited resolution under an excellent seeing condition, fully covering the M6.5 flare. The data taken include images in TiO (705.7~nm; 10~\AA\ bandpass) by the Broad-band Filter Imager with a field of view (FOV) of 70\arcsec\ at 0.1\arcsec\ resolution and 15~s cadence, Fabry-P{\'e}rot spectroscopic observations around the \ha\ line center at $\pm$1.0, $\pm$0.6, and 0.0~\AA\ (0.07~\AA\ bandpass) by VIS with a 70\arcsec\ circular FOV at 0.1\arcsec\ resolution and 28~s cadence, and spectropolarimetric observations of the Fe~{\sc i} 1564.8~nm line (0.1~\AA\ bandpass) by NIRIS with a 85\arcsec\ round FOV at 0.24\arcsec\ resolution and 87~s cadence (for a full set of Stokes measurement). Bursts of 100 and 25 frames are processed for speckle reconstruction at TiO and each \ha\ line position, respectively. In this study, we aligned \ha\ $+$~1.0~\AA\ images with sub-pixel precision and used these \ha\ far red-wing images to best trace the evolution of flare ribbon fronts \citep[e.g.,][]{deng13}. 
\begin{figure*}
\epsscale{1.17}
\plotone{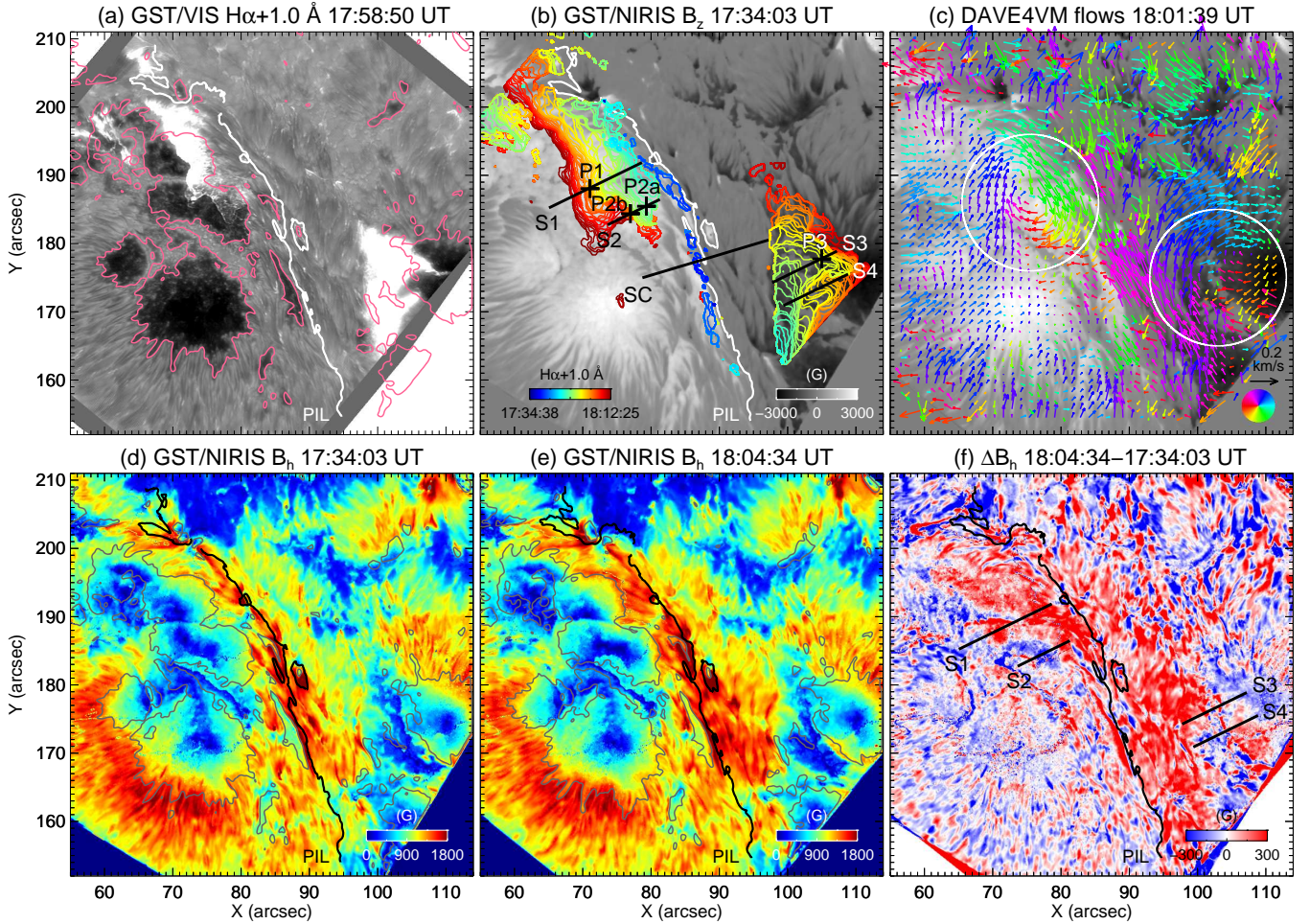}
\caption{Evolution of flare ribbons and magnetic fields. (a) \ha~$+$~1.0~\AA\ image near the flare peak showing the two major flare ribbons. The magenta lines contour $B_z$ map (smoothed by a window of 0.7\arcsec~$\times$~0.7\arcsec) at $\pm$1600~G. (b) $B_z$ image superimposed with curves (color-coded by time) that depict the progression of flare ribbon fronts. Note that the western ribbon and its evolution are not entirely captured due to the limited FOV of GST. The overplotted lines S1--S4 and SC indicate the slit positions of the time slices and vertical cross sections shown in Figures~\ref{f_slits} and \ref{f_vertical}, respectively; the magnetic field evolution in several sample positions (P1, P2a, P2b, and P3) is plotted in Figure~\ref{f_evo}. (c) $B_z$ image superimposed with arrows (color-coded by direction; see the color wheel) that illustrate DAVE4VM flows averaged between 17:52:56--18:13:17~UT (sunspot rotation phase; \citealt{liu16}) subtracted by the flow field averaged between 17:32:35--17:51:29~UT. The two white circles mark the regions of rotational motion. A window size of 23 pixels was set for DAVE4VM tracking. (d)--(f) Maps of $B_h$ in the pre- and postflare states and their difference. The PIL is overplotted in (a)--(b) and (d)--(f). All the images in this paper are aligned with respect to 17:34:03~UT. \edit1{An animation of the images in the panels (a), (b), and (f) is available. From left to right the sequences show images of \ha~$+$~1.0~\AA, $B_z$, and fixed difference of $B_h$ (relative to 17:34:03~UT). The sequences start at 2015 June 22 17:35:35~UT, 17:35:30~UT, and 17:35:30~UT, and end at 18:39:27~UT, 18:39:26~UT, and 18:39:26~UT, respectively. The video duration is 3~s.} \label{f_overview}}
\end{figure*}

It is notable that this M6.5 flare is one of the first major flare events observed by NIRIS, which is dedicated to the 1564.8~nm doublet band observation. This spectral line is the most Zeeman sensitive probe (with the maximum splitting factor Land{\'e} $g=3$) of the magnetic field within a small height range at the atmospheric minimum opacity, the deepest photosphere \citep{solanki92}, and is the best spectral line for umbral magnetic field observations in the entire electromagnetic spectrum \citep{harvey75,livingston15}. Although it has a lower diffraction limit than some visible lines and the issue of thermal noise has to be mitigated, the 1564.8~nm line has lower scattered light, produces more stable images under the circumstances of atmospheric turbulence, and only exhibits emissions in some extremely energetic flares. Equipped with two Fabry-P{\'e}rot etalons in a dual-beam optical design, NIRIS captures two simultaneous polarization states and images them side-by-side onto half of a closed-cycle, helium-cooled 2048~$\times$~2048 HgCdTe infrared array. Significant efforts have been devoted to develop the NIRIS data processing pipeline at BBSO \citep{ahn16,ahn17}, which essentially includes dark and flat field correction, image alignment and destretching for dual beams (with 60 wavelength sampling), calibration of instrumental crosstalk (by measuring the detector response to pure states of polarization passing through the telescope optics), and Stokes inversion using the Milne-Eddington (M-E) atmospheric approximation (with initial parameters pre-calculated to resemble the observed Stokes profiles). For a proper exploration of NIRIS vector field measurement, we further resolved the 180\dg\ azimuthal ambiguity using the ME0 code originally developed for \hinode\ vector data \citep{leka09b,leka09a} that is based on the ``minimum energy" algorithm \citep{metcalf94,metcalf06}, removed the projection effect by transforming the observed vector fields to heliographic coordinates \citep{gary_hagyard90}, and conducted a validation of data processing by comparing to HMI data products (see the Appendix and Figure~\ref{f_validation}). The NIRIS vector magnetograms deduced from the above procedures were used in our previous study of this event \citep{wang17}. Note that following the convention of \hinode, the disambiguated azimuth angle in this paper ranges counterclockwise from $-$180\dg\ to 180\dg, with the direction of zero azimuthal angle pointed to the solar west. In order to minimize the seeing effect (spatially varying image motion) in the ground-based observations, in this work we also performed image destretching to intensity images from the inversion, and then applied the determined destretch to the time sequence of NIRIS vector magnetograms. NIRIS intensity images were also used to accurately co-align NIRIS vector field observations with \ha\ far red-wing images through matching sunspot and plage areas.

\begin{figure*}
\epsscale{.805}
\plotone{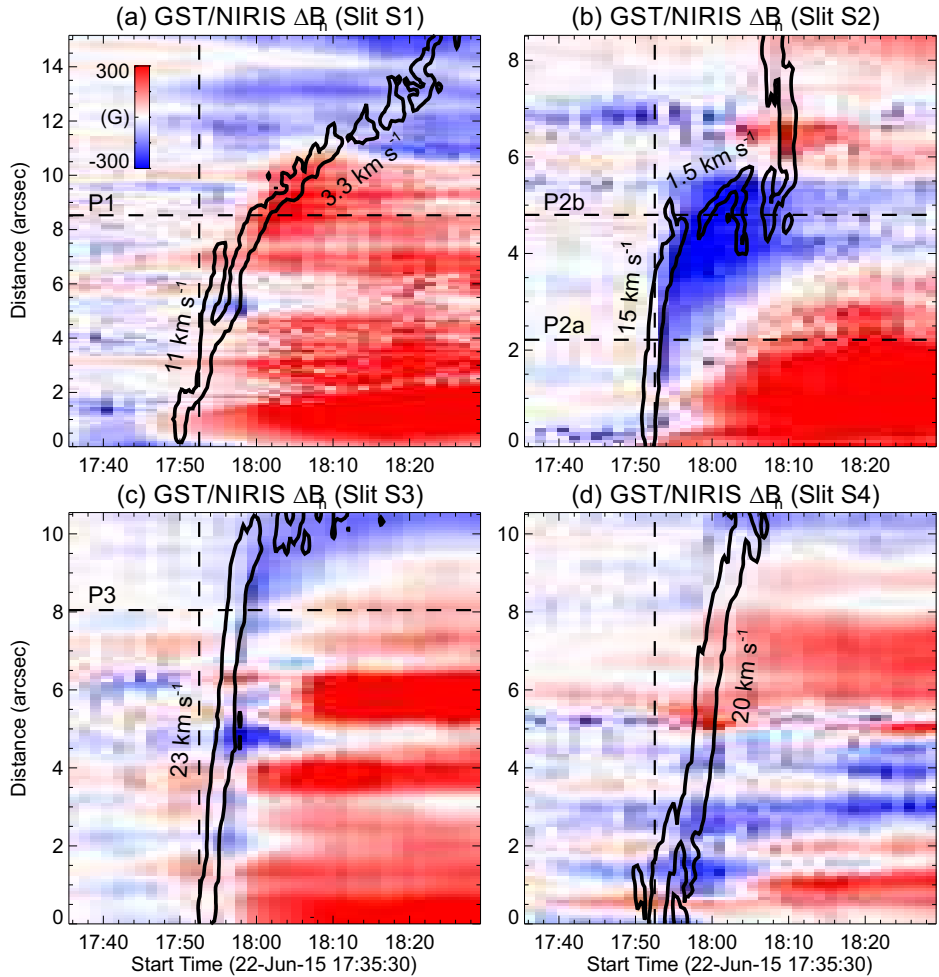}
\caption{Correlation between flare ribbon motion and $B_h$ change. The background portrays time slices for the slits S1--S4 (as denoted in Figure~\ref{f_overview}(b)) using the fixed difference images (relative to the preflare time at 17:34:03~UT) of $B_h$, showing the $B_h$ change. The superimposed black lines are contours (at 600 DN) of time slices for the slits S1--S4 (smoothed by a window of 0.23\arcsec~$\times$~0.23\arcsec) using the running difference \ha\ $+$~1.0~\AA\ images, showing the motion of ribbon front. The estimated ribbon velocities along each slit are denoted. In particular, along the slit S2, the speed of the ribbon front is \sm15~\kms\ during 17:52--17:55~UT and \sm1.5~\kms\ during 17:55--18:06~UT, as denoted in (b). For all the slits, the distance is measured from the end closest to the PIL. The horizontal dashed lines mark the positions of P1, P2a, P2b, and P3 relative to their corresponding slits. The vertical dashed line indicates the time of the first main HXR peak at 17:52:31~UT. \label{f_slits}}
\end{figure*}

The \sdo/HMI observations used to accompany the NIRIS data analysis are full-disk vector magnetograms at 1\arcsec\ resolution and 135~s cadence \citep{sun17}. The HMI instrument takes filtergrams of Stokes parameters at six wavelength positions around the Fe~{\sc i} 617.3~nm spectral line. The Stokes inversion technique implemented to routinely analyze HMI pipeline data is also based on the M-E approximation \citep{borrero11}, and a variant of the ME0 code is used for azimuthal disambiguation \citep{hoeksema14}. The retrieved HMI data were processed (mainly for combining disambiguation results with azimuth, and deprojection) using standard procedures in the Solar SoftWare (SSW) provided by the HMI team, and were expanded in size to match and align with NIRIS. For NLFFF extrapolations, we remapped HMI magnetograms of the entire AR at original resolution using Lambert (cylindrical equal area) projection centered on the middle point of the AR. After adjusting the photospheric boundary with a preprocessing procedure to better suit the force-free condition \citep{wiegelmann06}, we constructed a time sequence of NLFFF models using the ``weighted optimization'' method \citep{wheatland00,wiegelmann04} optimized for HMI data \citep{wiegelmann10,wiegelmann12}. The calculation was made using 2~$\times$~2 rebinned magnetograms within a box of 472 $\times$ 224 $\times$ 224 uniform grid points (corresponding to about 348 $\times$ 165 $\times$ 165 Mm$^3$). In addition, soft- and hard X-ray (HXR) emissions of the 2015 June 22 M6.5 flare were recorded by the \GOES\ (GOES)-15 and \textit{Fermi Gamma-Ray Burst Monitor} \citep{meegan09}, respectively. In GOES 1.6--12.4~keV energy flux, the flare of interest starts at 17:39~UT, peaks at 18:23~UT, and ended at 18:51~UT, with the first main peak in Fermi 25--50~keV HXR flux at 17:52:31~UT \citep{liu16}.

\begin{figure*}
\epsscale{1.07}
\plotone{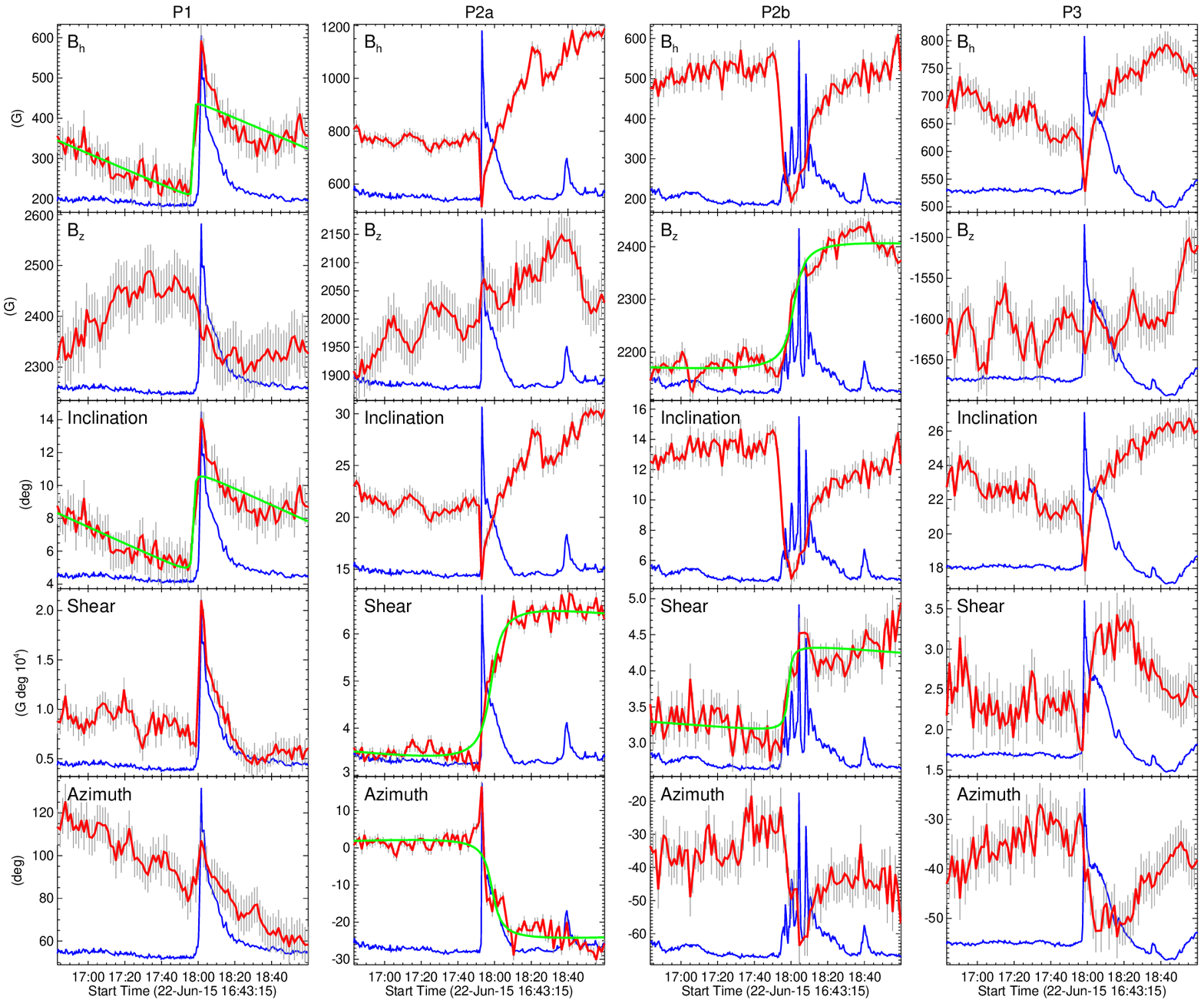}
\caption{Time profiles of flare \ha\ $+$~1.0~\AA\ emission and changes of photospheric magnetic field at sample positions P1, P2a, P2b, and P3 (as marked in Figure~\ref{f_overview}(b)). The \ha\ light curves are plotted in blue and in an arbitrary unit. The quantities plotted in red are, from top to bottom rows, $B_h$, $B_z$, inclination angle, magnetic shear, and azimuth angle. In each panel, the grey error bars indicate a $1\sigma$ level of the fluctuation of corresponding magnetic field parameter in the preflare time (from 16:43:15 to 17:38:24~UT). When appropriate, the field evolution is fitted using a step function (green lines). \label{f_evo}}
\end{figure*}

\section{ANALYSES AND RESULTS} \label{results}
Figure~\ref{f_overview} presents an overview of the evolution of chromospheric ribbons and photospheric field in the 2015 June 22 M6.5 flare. Here the FOV of BBSO/GST covers the central core region of the flare. From animations of VIS \ha\ $+$~1.0~\AA\ and NIRIS $B_z$ (available in the online journal), it can be clearly seen that (1) two main flare ribbons move away from the PIL and sweep through two sunspot regions of opposite polarities (also see Figures~\ref{f_overview}(a) and (b)), and (2) both sunspots undergo a clockwise rotation during the flare period, which is unambiguously demonstrated with flow tracking using the differential affine velocity estimator for vector magnetograms \citep[DAVE4VM;][]{schuck08} method (see Figure~\ref{f_overview}(c)). This is consistent with previous studies using TiO and HMI observations \citep{liu16,bi17}. Interestingly, the southern part of the eastern ribbon apparently slows down when approaching the center of the eastern rotating sunspot (cf. Figures~\ref{f_overview}(b) and (c)). A similar but less obvious slowdown is discernible for the central part of the western ribbon. A comparison between pre- and postflare images (see Figures~\ref{f_overview}(d) and (e), and also the $B_h$ animation) shows that there is a pronounced enhancement of $B_h$ in an extended region mainly along the PIL (red-colored region in Figure~\ref{f_overview}(f)). To better disclose the $B_h$ evolution, we make fixed difference $B_h$ images relative to a preflare time. From the time-lapse movie, it is remarkable to notice that the enhancement of $B_h$ not only shows up around the PIL \citep{wangjs18}, but also moves away from the PIL and spreads across the flaring region, mimicking the flare ribbon motion. More intriguingly, a negative $\delta B_h$ front, meaning a transient weakening of $B_h$, appears to precede the moving $B_h$ enhancement, especially at the southern portion of the eastern ribbon and the entire western ribbon.

To accurately characterize the $B_h$ evolution associated with the flare ribbon motion, time-distance maps along slits S1--S4 (drawn in Figure~\ref{f_overview}(b)) based on the fixed difference $B_h$ images are presented as the backgrounds of Figure~\ref{f_slits}. They are overplotted with contours of the same time-distance maps but based on the running difference \ha\ $+$~1.0~\AA\ images that highlight the ribbon fronts. We constructed these slits by orientating elongated windows (with various length but a common short side of 0.78\arcsec) approximately perpendicular to the observed ribbon motion at 26\dg\ counterclockwise from the solar west, and averaged the pixels across the short sides. The distance shown is measured from the ends of slits closest to the PIL. In Figure~\ref{f_evo}, the temporal evolution of \ha\ $+$~1.0~\AA\ emission (blue) is compared with that of vector magnetic field (red; in terms of $B_h$, $B_z$, inclination angle relative to the vertical direction, magnetic shear, and azimuth angle) at several representative positions P1, P2a, P2b, and P3 along the slits (as marked in Figure~\ref{f_overview}(b); values averaged over 7~$\times$~7 pixels$^2$ centered on them). Here the magnetic shear for evaluating the nonpotentiality is computed as $B \cdot \theta$ \citep{wang94,wang06shear}, where $B=|\mathbi{B}|$ and $\theta={\rm cos}^{-1}(\mathbi{B} \cdot \mathbi{B}_p)/(B\,B_p)$, with the subscript $p$ representing the potential field, which we derived using the fast-Fourier transform method \citep{alissandrakis81}. When appropriate, we also fit these time profiles of magnetic properties with a step function \citep[green lines;][]{sudol05}. Based on these results, we observe the \edit1{following}.

Along the slit S1, there exists a close spatial and temporal correlation between the motion of the eastern flare ribbon and the enhancement of $B_h$ (Figure~\ref{f_slits}(a)), especially after the time of the first main HXR peak (vertical dashed line). At P1 (see Figure~\ref{f_evo}, first column), with the arrival of ribbon front the photospheric field turns more inclined relative to the surface, with $B_h$ and inclination angle increased stepwise by 244$\pm$24~G and 6.4$\pm$0.6\dg\ in \sm0.5 and 1.5 minutes, respectively; also, magnetic shear sharply increases by \sm250\% but then returns to the preflare level in about 20 minutes. In contrast, $B_z$ evolved more gradually without an abrupt change. In the meantime, a transient increase of azimuth angle meaning a temporary counterclockwise rotation of field vectors can be noticed \citep{xu18}.

\begin{figure*}
\epsscale{1.17}
\plotone{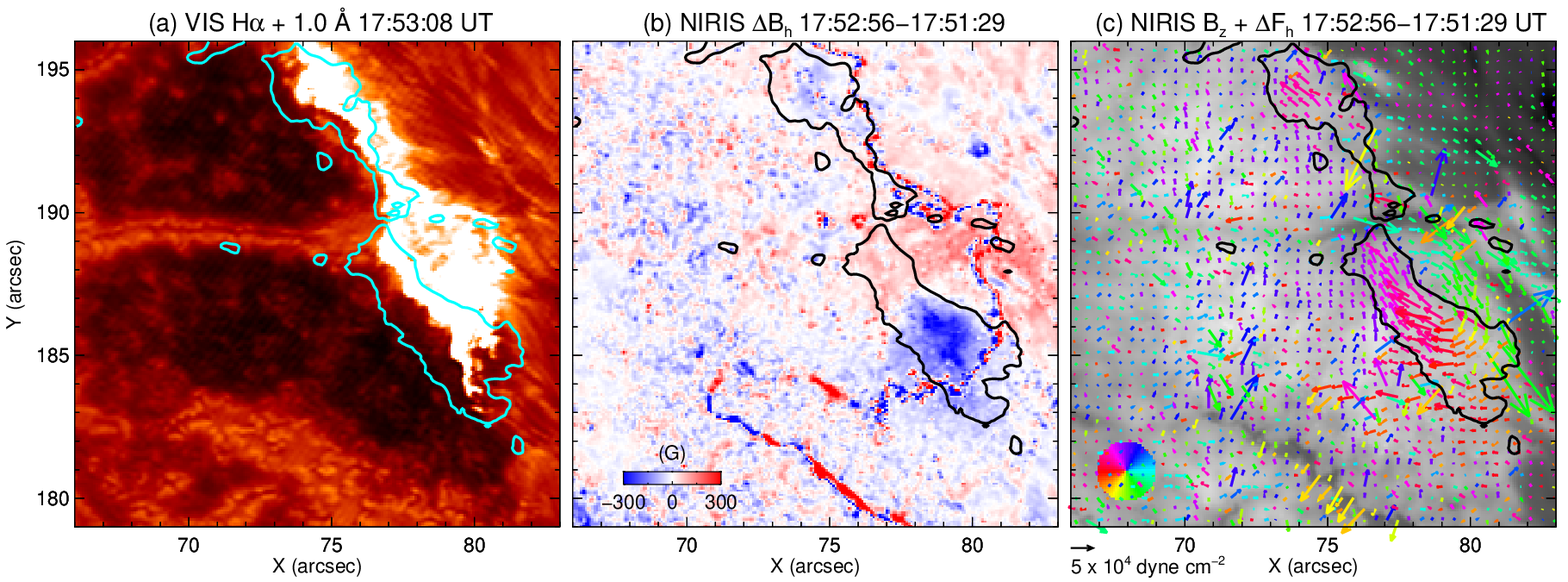}
\caption{Flare ribbon and induced Lorentz-force change. (a) \ha~$+$~1.0~\AA\ image at 17:53:08~UT near the first main HXR peak, overplotted with contours (smoothed by a window of 0.55\arcsec~$\times$~0.55\arcsec) at 600 DN (same level as that used in Figure~\ref{f_slits}) based on the running difference \ha~$+$~1.0~\AA\ image (i.e., 17:53:08 minus 17:52:40~UT) that highlight the ribbon front. (b) Running difference image of $B_h$ at about the same time. (c) The corresponding $B_z$ image (scaled from $-$1000 to 3000~G) overplotted with arrows (color-coded by direction; see the color wheel) representing $\delta F_h$ vectors. The contours in (b) and (c) are the same as those plotted in (a). \label{f_lorentz}}
\end{figure*}

Along the slit S2 across the center of the eastern rotating sunspot, the propagation of the eastern flare ribbon exhibits a prominent deceleration, and the arrival of the ribbon front is coincident with a transient decrease of $B_h$ followed by an increase (see Figure~\ref{f_slits}(b)). At P2a (see Figure~\ref{f_evo}, second column), $B_h$ and inclination angle temporarily decrease by \sm300~G and \sm8\dg\ and then increases by \sm600~G and \sm13\dg\ in \sm30 minutes, respectively; meanwhile, magnetic shear shows a step-like increase by \sm86\% in \sm15 minutes. After a transient increase like at P1, azimuth angle begins to decrease, connoting the observed clockwise sunspot rotation \citep{liu16} that drags the magnetic field with it. Compared to P2a, the magnetic field evolution at P2b (around the rotation center) bears a resemblance but displays a more prolonged decrease of $B_h$ and inclination angle; remarkably, $B_z$ at P2b undergoes a permanent increase of 266$\pm$20~G in \sm13 minutes around 18~UT (see Figure~\ref{f_evo}, third column), when the speed of the flare ribbon has evidently reduced (Figure~\ref{f_slits}(b)). Since darker umbrae evince stronger vertical fields \citep{martinez93}, the irreversible increase of $B_z$ of this rotating sunspot is also evidenced by a \sm7\% decrease of its overall intensity in TiO and 1564.8~nm after the flare \citep{liu16,deng17}. \edit1{A line profile analysis further corroborates that the transient decrease of $B_h$ (and also increase of $B_z$) at P2b is irrelevant to magnetic anomaly due to flare heating (see the Appendix and Figure~\ref{stokes})}. Assuming a uniform reconnecting electric field along the entire eastern ribbon, the observed slowdown of flare ribbon motion with concurrent increase of $B_z$ at a portion of the ribbon could be expected \citep{forbes84}. From about 18:08~UT, the northern section of the ribbon curves southward and overtakes the motion of the ribbon along S2.

In Figure~\ref{f_lorentz}, we further compare the locations of the eastern flare ribbon front, $\delta B_h$, and the horizontal Lorentz-force change $\delta {\mathbf F_{\rm h}} = \frac{1}{4\pi}\int dA \delta (B_{\rm r}{\mathbf B_{\rm h}})$ \citep{fisher12}, at a time close to the first main HXR peak. Note that the newly brightened ribbon region (e.g., the ribbon front) is cospatial with the region of decreased $B_h$. This, together with the increased $B_h$ at the region just swept by the ribbon, yields a vortex pattern in the $\delta {\mathbf F_{\rm h}}$ map. Obviously, the torque provided by this $\delta {\mathbf F_{\rm h}}$ vortex has the same direction (i.e., clockwise) as the observed sunspot rotation. This implies that the $B_h$ decrease preceding its increase may create a moving horizontal Lorentz-force change to drive the differential sunspot rotation as observed \citep{liu16}.

Along the slit S3, the motion of the western ribbon is correlated with magnetic field changes in a way similar to those found along the slit S2, e.g., showing a transient decrease followed by an increase of $B_h$ and inclination angle (see Figure~\ref{f_slits}(c) and light curves of P3 in the fourth column of Figure~\ref{f_evo}). A ribbon deceleration together with a prolonged decrease of $B_h$ also seems to be present along the slit S4 across the center of the western rotating sunspot (see Figure~\ref{f_slits}(d)). It is worthwhile to mention that despite of a lower resolution, vector magnetograms from HMI show very similar magnetic field changes related to flare ribbon motions as described above (see the Appendix and Figure~\ref{f_hmi_slits}), which substantiates the NIRIS results.

\begin{figure*}
\epsscale{1.08}
\plotone{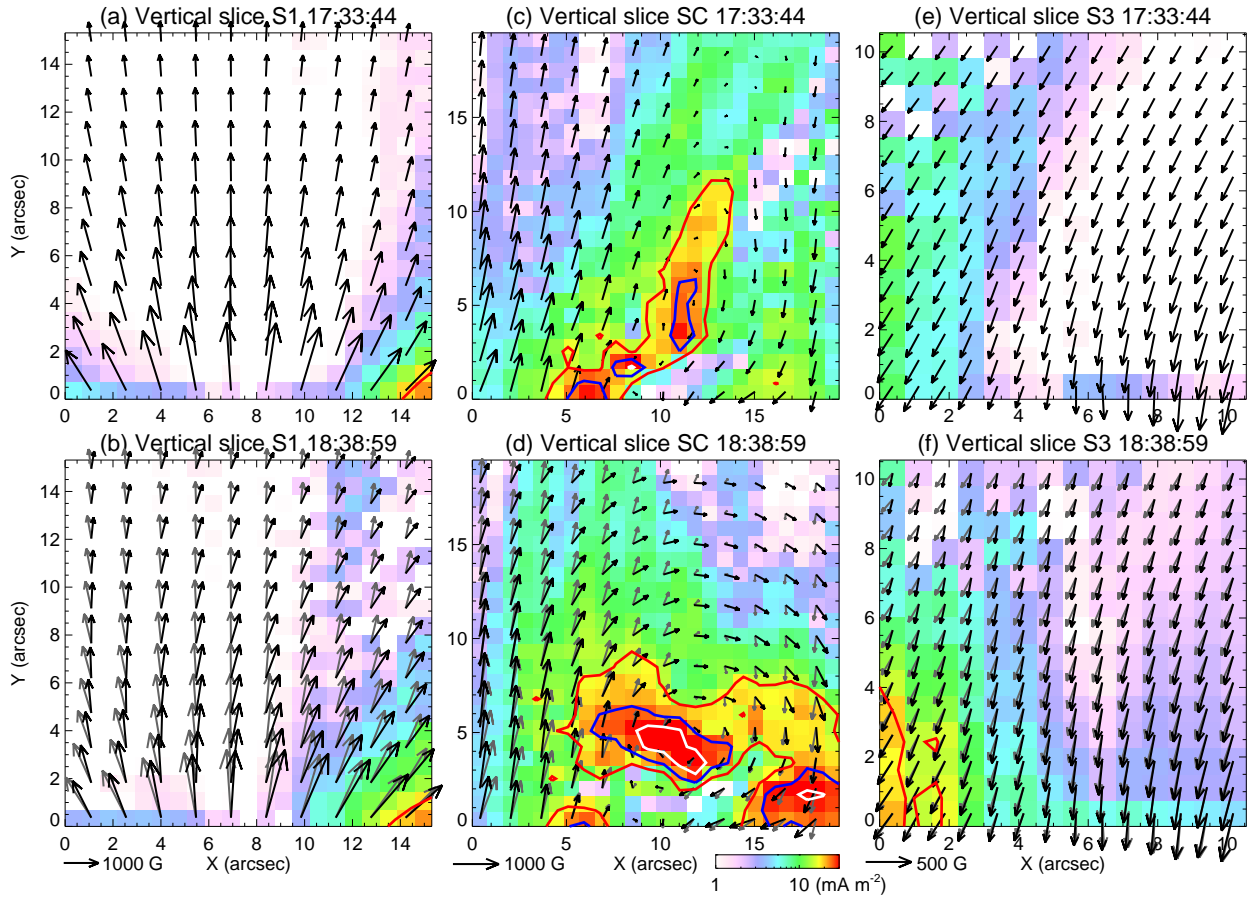}
\caption{Distributions of magnetic field and current in the preflare (a, c, e) and postflare (b, d, f) states in vertical cross sections S1, SC, and S3, the bottom sides of which are slits S1, SC, and S3, respectively, as denoted in Figure~\ref{f_overview}(b). The distance on the surface is measured from east to west for all slices. The background shows $J_h$ in logarithmic scale, overplotted with black arrows representing the transverse field vectors in the vertical slices. The preflare field vectors are also shown in gray in the corresponding postflare maps. The red, blue, and white contours are at levels of 0.015, 0.023, and 0.031 A~m$^{-2}$, respectively. \edit1{An animation showing the evolution from what is displayed in the top row to the bottom row is available. The sequences start at 2015 June 22 17:33:44~UT and end at 18:38:59~UT. The video duration is 3~s.} \label{f_vertical}}
\end{figure*}

Finally, we investigate the flare-related coronal field evolution in terms of the distribution of horizontal component of the total electric current density $|J_h|=(J_x^2+J_y^2)^{1/2}$ in several vertical slices to the extrapolated 3D coronal magnetic field. These slices intersect with the surface at locations of the same slits S1 and S3 as above through the regions of flare ribbons and another slit SC perpendicularly across the central PIL (as denoted in Figure~\ref{f_overview}(b)). Plotted in the top and bottom rows of Figure~\ref{f_vertical} are the distributions of $J_h$ for the pre- and postflare states, respectively, in these vertical slices, which are superimposed with arrows representing the transverse magnetic field vectors. From the results and also the supplementary animation spanning the flaring period, it transpires that a downward collapse of coronal field occurs intimately associated with the flare \edit1{\citep[e.g.,][]{sun12,liu12,liu14}}. This is visualized by the dramatic change of the coronal currents above the PIL, from a vertically elongated source reaching 12\arcsec\ to a substantially enhanced, horizontally elongated source concentrated close to the surface below 9\arcsec\ (cf. Figures~\ref{f_vertical}(c) and (d)). \edit1{We further show that} the collapse is also manifested by the clockwise (counterclockwise) turning of magnetic field vectors in the east (west) side of the PIL (except that the near-surface region in S3 has a clockwise turning), which leads to a more horizontal (i.e., inclined) configuration of magnetic fields at and above regions of the PIL and flare ribbons, conforming to the observed surface $B_h$ enhancement therein. We note that (1) field vectors in the far east portion of S1 (and also the upper portion of S3) become more vertical after the flare. This reflects the fact that in the outer flaring region, $B_h$ is observed to decrease  (see Figures~\ref{f_overview}(f) and \ref{f_slits}(a)) together with weakened penumbral features (not shown), which may be coherent with the collapse of the central fields \citep[e.g.,][]{liu05}. (2) Although not well demonstrated by the present extrapolations, it is plausible to expect that a time sequence of coronal field models with higher spatial and temporal resolution might show the successive turning of field vectors with the motion of flare ribbons.

\section{SUMMARY AND DISCUSSIONS} \label{summary}
In this paper, we take advantage of BBSO/GST high-resolution observations of both chromospheric ribbons in VIS off-band \ha\ and NIRIS photospheric vector magnetic fields in near infrared during the 2015 June 22 M6.5 flare to carry out a detailed investigation of photospheric vector magnetic field changes with related to flare ribbon motions, \edit1{which were not studied before}. This large and complex event shows not only the separation of flare ribbons but also the flare-related rotations of sunspots. We analyzed the permanent surface magnetic field changes in the flare ribbon regions with a focus on $B_h$, using time-distance maps and temporal evolution plots. We also explored the 3D coronal restructuring with aid from the NLFFF modeling based on \sdo/HMI vector magnetograms. Major findings are summarized as follows.

\begin{enumerate} 

\item In the photosphere, $B_h$ increases with the flare occurrence and this enhancement propagates away from the central PIL across the flaring region, exhibiting a close spatial and temporal correlation with the flare ribbon motion especially after the first main HXR peak (Figure~\ref{f_slits}). As seen in several representative positions (Figure~\ref{f_evo}), the strengthening of $B_h$ (by \sm300~G) at the arrival of the flare ribbon front is accompanied by an increase of inclination angle (by \sm6\dg), indicating that magnetic field becomes more inclined to the surface; also, the nonpotentiality as represented by magnetic shear generally enhances.

\item At the locations where azimuth angle sharply decreases indicating the sudden sunspot rotation, $B_h$ and inclination angle decrease transiently before being enhanced. Particularly, the flare ribbon decelerates toward the sunspot rotation center where $B_z$ becomes greatly \edit1{intensified} (Figures~\ref{f_slits} and \ref{f_evo}).

\item In the corona, a downward collapse of coronal magnetic field by \sm3\arcsec\ toward the photosphere is clearly portrayed by the evolution of the vertical profiles of $J_h$ around the PIL \edit1{\citep{sun12,liu12,liu14}}, which changes from a vertically elongated source to an enhanced, horizontally elongated source close to the surface (Figure~\ref{f_vertical}). Correspondingly, above the PIL and flare ribbon regions, magnetic field becomes more inclined, which is consistent with the observed enhancement of $B_h$. We surmise that a successive turning of field vectors associated with the flare ribbon motion might be visualized given coronal field models with a sufficiently high resolution.

\end{enumerate}

The increase of $B_h$ at flaring PILs between flare ribbons has been known from the previous observations \citep[e.g.,][]{sun12,liu12,wangs12}. The distinctive finding made in this investigation is that $B_h$ enhances not only at the PIL region, but at the locations of the flare ribbon fronts. As the flare ribbons move away from the PIL, such enhancements also propagate successively with the ribbons. This discovery of the flare-ribbon-related photospheric field changes could be made owing to the high resolution of NIRIS observations, and is also substantiated by the HMI data. Since it has been well established that flare ribbon fronts are the footpoints of the newly reconnected field lines in the corona, the vector field changes spatiotemporally correlated with the ribbon fronts must be a nearly instantaneous response of photospheric fields to the coronal restructuring, specifically, the reconnection of individual flux bundles.

We also want to point out that the correlation between the eastern flare ribbon and its related vector field change is complicated by the fact that ahead of the eastern ribbon, there is another elongated small brightening that propagates from north to south (see the movie), along a line of high values of the squashing factor $Q$ (\citealt{titov02}; see the Appendix and Figure~\ref{f_slit5}(a)). The high-$Q$ lines correspond to the footprints of quasi-separatrix layers \citep{demoulin96,demoulin97}, which are known to be favorable positions of flare ribbons. This brightening joins with the main eastern ribbon in the north (out of the FOV of GST) to form a continuous ribbon structure. To check whether this extra flare ribbon introduces magnetic field changes, we place a slit S5 perpendicular to the northern portion of the eastern ribbon (Figure~\ref{f_slit5}(a)) and repeat the analysis as done in Figure~\ref{f_slits}. Both the results using NIRIS and HMI data evince that the enhancement of $B_h$ not only appears to follow the movement of the main eastern ribbon, but also occurs ahead of it, distending to the region of the extra ribbon (see Figures~\ref{f_slit5}(b) and (c)). We consider this as an additional piece of evidence that the photospheric vector magnetic field may respond nearly instantaneously to the coronal reconnection.

There are a few models that may help understanding the present observations. The series of force-free field models give only snapshots of equilibrium states rather than dynamic evolution; nevertheless, the disclosed redistribution of electric current system may reflect a coronal field restructuring following magnetic energy release in the corona \citep[e.g.,][]{hudson00}. A back reaction of such coronal magnetic reconfigurations on the photosphere and interior may be expected \citep{hudson08}, but it only loosely points to a more horizontal photospheric field, i.e., an increase of $B_h$; further, it does not necessarily explain why the magnetic shear should also increase. The shear Alfv{\'e}n wave model \citep{wheatland18} can explain both the increase of $B_h$ and magnetic shear, in which the shear Alfv{\'e}n waves launched from the coronal reconnection region travel downward to impact the flare ribbon regions. In 3D, these waves correspond to the torsional Alfv{\'e}n waves so that the rotation of plasma and magnetic field at the ribbon location is also expected. In addition, we have presented an idea that the $B_h$ decrease preceding its increase may create a moving horizontal Lorentz-force change (Figure~\ref{f_lorentz}) to drive the differential sunspot rotation as observed \citep{liu16}. It remains puzzling why $B_h$ decreases at the region of the newly brightened ribbon.

Our main intention of this study is to present the details of the new phenomenon of the flare-ribbon-related photospheric magnetic field changes. It remains to see whether these vector field changes as found in this event are a generic feature of all flares or simply a peculiarity of this event. Certainly, more simultaneous high-resolution observations of chromospheric flare ribbons and photospheric vector magnetic fields throughout the flaring period are much desirable to further elucidate the photosphere-corona coupling in the flare-related phenomena.

\acknowledgments
We thank the teams of BBSO and \sdo\ for providing the observational data of this event. The BBSO operation is supported by NJIT and US NSF AGS 1821294 grant. The GST operation is partly supported by the Korea Astronomy and Space Science Institute and Seoul National University, and by the strategic priority research program of \edit1{Chinese Academy of Science (CAS)} with grant No. XDB09000000. C.L., N.D., and H.W. were supported by NASA grants NNX13AF76G, NNX13AG13G, NNX16AF72G, 80NSSC17K0016, 80NSSC18K0673, and 80NSSC18K1705, and by NSF grants AGS 1408703 and 1821294. W.C. was supported by the grants of NSF-AGS 1821294 and \edit1{National Science Foundation of China (NSFC)} 11729301. J.C. was supported by the Korea Astronomy and Space Science Institute under the R\&D program, Development of a Solar Coronagraph on International Space Station (Project No. 2017-1-851-00), supervised by the Ministry of Science, \edit1{Information and Communications Technology (ICT)}, and Future Planning. D.P.C. was supported by NSF grants AGS 1413686 and 1620647. J.L. was supported by NSFC grants 41331068, 11790303 (11790300), and 41774180. R.L. was supported by NSFC grants 41474151, 41774150, and 41761134088.

\vspace{5mm}
\facilities{BBSO/GST, \sdo(HMI)}

\appendix

\section{Appendix information}
\begin{figure*}[!t]
\epsscale{1.17}
\plotone{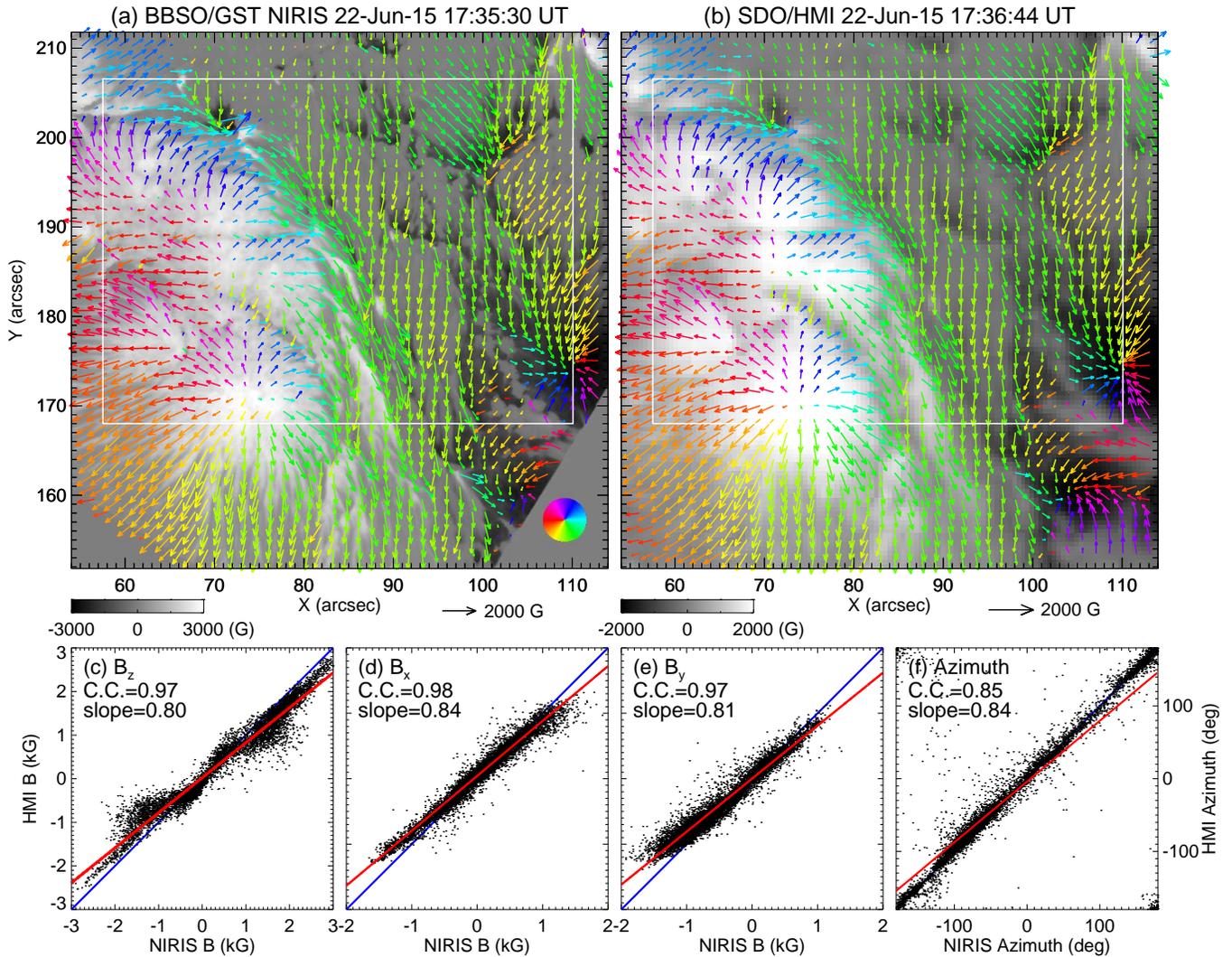}
\caption{Comparison between BBSO/GST NIRIS and \sdo/HMI vector magnetograms of NOAA AR 12371. The data were taken at about the same time and processed in a similar fashion (Stokes inversion, azimuth disambiguation, and deprojection; see text for details). (a)--(b) Images of $B_z$ superimposed with arrows (color-coded by direction; see the color wheel) representing vectors of $B_h$. (c)--(f) Scatter plots of NIRIS vs. HMI measurements of $B_z$, $B_x$, $B_y$, and azimuth angle for the boxed region marked in the upper panels (the higher resolution NIRIS images are downsampled by a factor of 6.4). Also indicated are the linear Pearson correlation coefficient (C.C.) and slope of linear fit of the data points (red lines). The underlying blue lines have a slope of 1. \label{f_validation}}
\end{figure*}

As a validation of GST/NIRIS data processing procedures, in Figure~\ref{f_validation} we compare vector magnetograms of NOAA AR 12371 from NIRIS and \sdo/HMI obtained at about the same time right before the 2015 June 22 M6.5 flare. The HMI data used is the full-disk vector magnetic field product (\verb|hmi.B_720s|), processed using standard procedures in SSW. It is clear that for the flare core region (white box in Figures~\ref{f_validation}(a) and (b)), both $B_z$ and $B_h$ field vectors derived from NIRIS and HMI measurements have a high correlation (see Figures~\ref{f_validation}(c)--(f)). The slope of \sm0.8 shown by the scatter plots indicates that NIRIS tends to produce stronger fields, presumably due to the fact that NIRIS observes at a deeper atmosphere than HMI.

\begin{figure*}[!t]
\epsscale{1.17}
\plotone{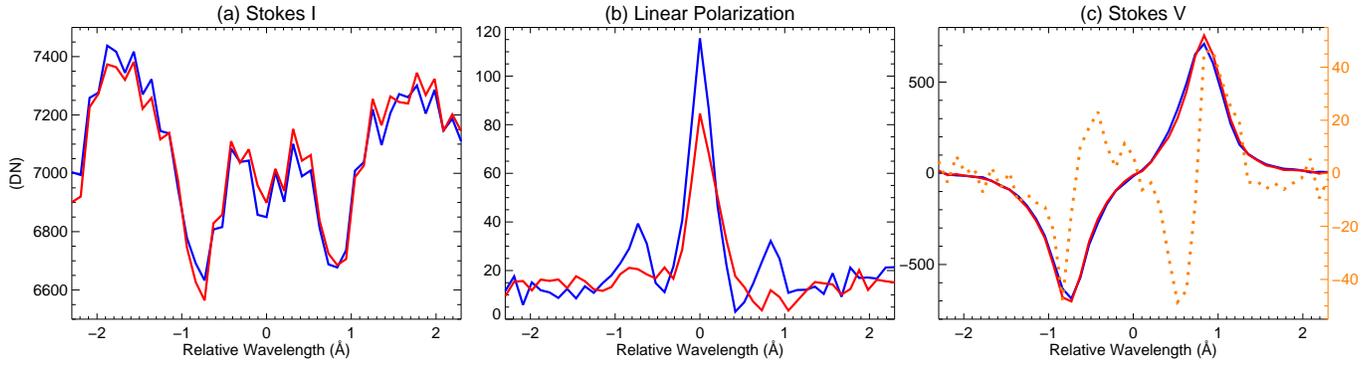}
\caption{\edit1{Changes of 1564.8~nm Stokes profiles at P2b associated with the arrival of flare ribbon. The profiles of Stokes $I$ (a), overall linear polarization magnitude $(Q^2+U^2)^{1/2}$ (b), and Stokes $V$ (c) at 17:34:03~UT (blue) and 18:04:34~UT (red) are plotted. In (c), the orange dotted line shows the difference profile (the profile at 18:04:34 UT is subtracted by that at 17:34:03 UT).} \label{stokes}}
\end{figure*}

\edit1{In Figure~\ref{stokes}, we present 1564.8~nm Stokes profiles at P2b before and after the arrival of the flare ribbon, at 17:34:03~UT and 18:04:34~UT, respectively. Comparing the results, we see that the Stokes $I$ component shows no clear and systematic changes (Figure~\ref{stokes}(a)), suggesting that flare heating does not alter the spectral line profiles. In contrast, the Stokes $QU$ combination $(Q^2+U^2)^{1/2}$ that measures the overall linear polarization magnitude \citep[e.g.,][]{leka01,deng10} obviously weakens (Figure~\ref{stokes}(b)), while the Stokes $V$ component representing the circular polarization enhances (see Figure~\ref{stokes}(c) and note the difference profile in orange). As this AR is close to the disk center at the time of the M6.5 flare, these changes of Stokes $QUV$ profiles are consistent with the observed decrease (increase) of the horizontal (vertical) field at this location, as presented in Figure~\ref{f_evo} (third column).} 

\begin{figure*}[!t]
\epsscale{0.805}
\plotone{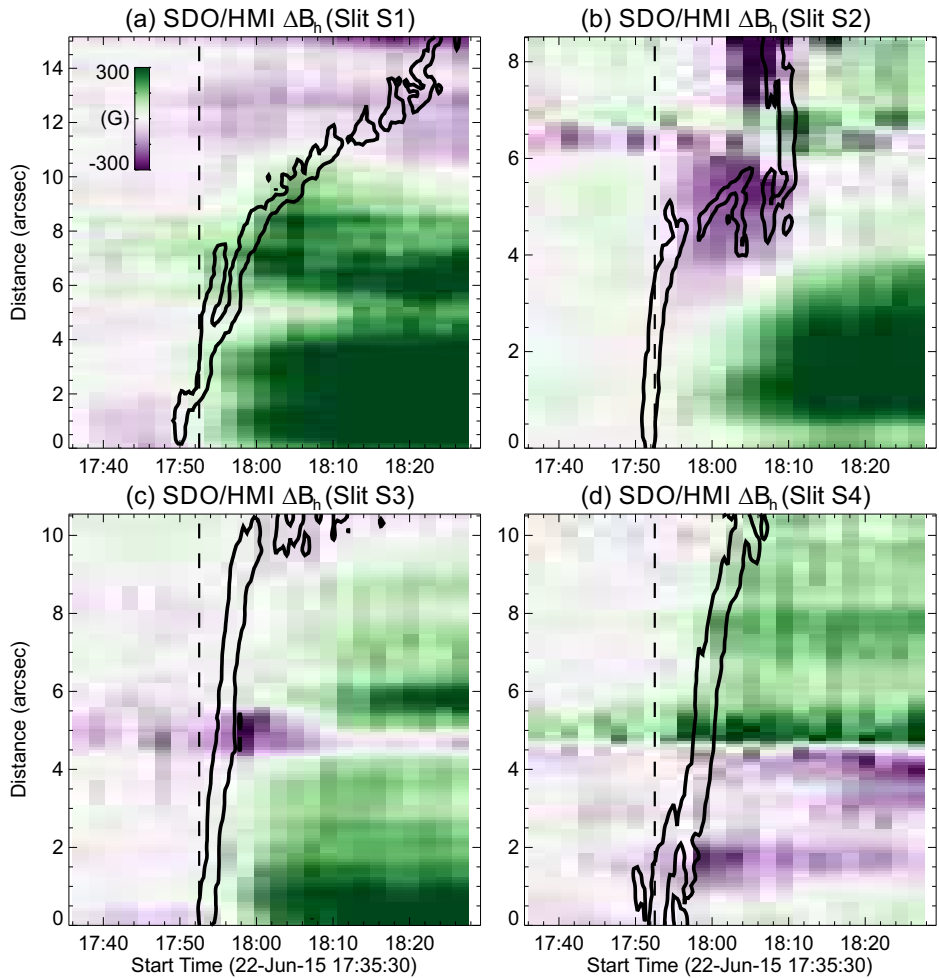}
\caption{Same as Figure~\ref{f_slits} but the background shows time slices for the slits S1--S4 using the fixed difference images of \sdo/HMI $B_h$. \label{f_hmi_slits}}
\end{figure*}

The presented analyses applied to NIRIS data were also carried out using HMI vector magnetograms, and generally similar results were obtained. In Figure~\ref{f_hmi_slits}, we show the time-distance maps along the slits S1--S4 based on the 135~s cadence HMI data. The results, despite of having a lower resolution, show evolutionary patterns that are almost identical to those obtained using the NIRIS data (see Figure~\ref{f_slits}).

\begin{figure*}[!t]
\epsscale{1.17}
\plotone{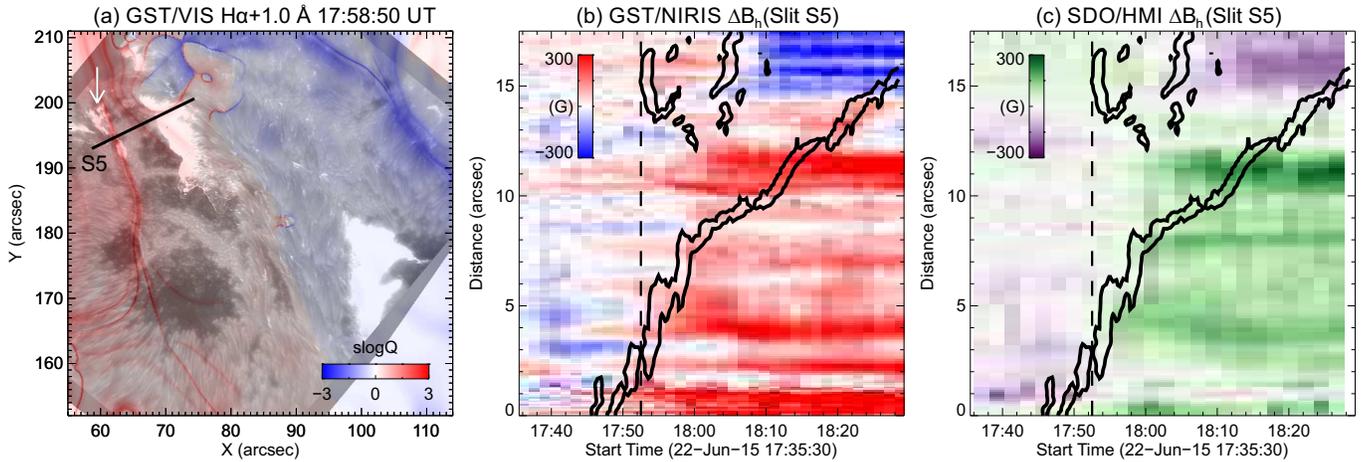}
\caption{Evolution of flare ribbon and magnetic field. (a) \ha~$+$~1.0~\AA\ image near the flare peak blended with the slog$Q$ map calculated at a preflare time at 17:36:44~UT, showing that in front of the major eastern flare ribbon, there is elongated emission as pointed to by the white arrow that is located at the high-$Q$ line. Panels (b) and (c) are similar to those shown in Figures~\ref{f_slits} and \ref{f_hmi_slits}, respectively, but for the slit S5 marked in (a). \label{f_slit5}}
\end{figure*}

In Figure~\ref{f_slit5}(a), an \ha~$+$~1.0~\AA\ image is blended with the derived map of slog$Q$, which is defined as ${\rm slog}Q = {\rm sign}(B_z){\rm log_{10}}Q$ \citep{titov11}. The calculation was conducted based on the potential field within the same box volume as the NLFFF, with the code developed by \citet{liur16}.

\end{document}